\documentclass{emulateapj}

\usepackage{amssymb}

\def\ts     {\thinspace}
\def\kms    {\ts km\ts s$^{-1}$}
\def\etal   {{\rm et\ts al.}}
\def\msol   {$M_{\odot}$}
\def\msolyr {$M_{\odot}$\,yr$^{-1}$}

\def\lprime {K\,\ts km\ts s$^{-1}$\,pc$^2$}

\def\aco    {{\rm CO}($J$=1$\to$0)}

\def\cco    {{\rm CO}($J$=3$\to$2)}
\def\ccoalt {{\rm CO} $J$=3$\to$2}

\def\eco    {{\rm CO}($J$=5$\to$4)}

\slugcomment{draft version \today, accepted for publication in the Astrophysical Journal Letters}

\shorttitle{CO($J$=1$\to$0) and CO($J$=5$\to$4) Imaging of a $z$=2.5 Submillimeter Galaxy}
\shortauthors{Riechers et al.}

\begin{document}

\title{Imaging the Molecular Gas Properties of a Major Merger \\ Driving the Evolution of a $z$=2.5 Submillimeter Galaxy}

\author{Dominik A.\ Riechers\altaffilmark{1,8}, Christopher L.\ Carilli\altaffilmark{2}, Fabian Walter\altaffilmark{3}, Axel Weiss\altaffilmark{4}, \\ Jeff Wagg\altaffilmark{5}, Frank Bertoldi\altaffilmark{6}, Dennis Downes\altaffilmark{7}, Christian Henkel\altaffilmark{4}, and Jacqueline Hodge\altaffilmark{3}}

\altaffiltext{1}{Astronomy Department, California Institute of
  Technology, MC 249-17, 1200 East California Boulevard, Pasadena, CA
  91125, USA; dr@caltech.edu}

\altaffiltext{2}{National Radio Astronomy Observatory, PO Box O, Socorro, NM 87801, USA}

\altaffiltext{3}{Max-Planck-Institut f\"ur Astronomie, K\"onigstuhl 17, D-69117 Heidelberg, Germany}

\altaffiltext{4}{Max-Planck-Institut f\"ur Radioastronomie, Auf dem H\"ugel 69, Bonn, D-53121, Germany}

\altaffiltext{5}{European Southern Observatory, Alonso de C{\'o}rdova 3107, Vitacura, Casilla 19001, Santiago 19, Chile}

\altaffiltext{6}{Argelander-Institut f\"ur Astronomie, Universit\"at Bonn, Auf dem H\"ugel 71, Bonn, D-53121, Germany}

\altaffiltext{7}{Institut de RadioAstronomie Millim\'etrique, 300 Rue de la Piscine, Domaine Universitaire, 38406 Saint Martin d'H\'eres, France}

\altaffiltext{8}{Hubble Fellow}


\begin{abstract}

We report the detection of spatially extended \aco\ and \eco\ emission
in the $z$=2.49 submillimeter galaxy (SMG) J123707+6214, using the
Expanded Very Large Array and the Plateau de Bure Interferometer. The
large molecular gas reservoir is spatially resolved into two \aco\
components (north-east and south-west; previously identified in
\ccoalt\ emission) with gas masses of 4.3 and 
3.5$\times$10$^{10}\,(\alpha_{\rm CO}/0.8)$\,\msol.  We thus find that
the optically invisible north-east component slightly dominates the
gas mass in this system. The total molecular gas mass derived from the
\aco\ observations is $\gtrsim$2.5$\times$ larger than estimated from
\cco.  The two components are at approximately the same redshift, but
separated by $\sim$20\,kpc in projection.  The morphology is
consistent with that of an early-stage merger.  The total amount of
molecular gas is sufficient to maintain the intense 500\,\msolyr\
starburst in this system for at least $\sim$160\,Myr.  We derive line
brightness temperature ratios of $r_{31}$=0.39$\pm$0.09 and
0.37$\pm$0.10, and $r_{51}$=0.26$\pm$0.07 and 0.25$\pm$0.08 in the two
components, respectively, suggesting that the $J$$\geq$3 lines are
substantially subthermally excited. This also suggests comparable
conditions for star formation in both components. Given the similar
gas masses of both components, this is consistent with the comparable
starburst strengths observed in the radio continuum emission.  Our
findings are consistent with other recent studies that find evidence
for lower CO excitation in SMGs than in high-$z$ quasar host galaxies
with comparable gas masses. This may provide supporting evidence that
both populations correspond to different evolutionary stages in the
formation of massive galaxies.

\end{abstract}


\section{Introduction}

Detailed studies of submillimeter galaxies (SMGs; see review by Blain
\etal\ \citeyear{bla02}) have revealed that they represent a relatively 
rare, but cosmologically important high redshift population of massive
galaxies. They harbor intense ($>$500\,\msol\,yr$^{-1}$), often
heavily obscured, but rather short-lived ($<$100\,Myr) starbursts that
rapidly consume their gas content through star formation at high
efficiencies.  SMGs may trace a common phase in the formation and
evolution of massive galaxies in the early universe, making them the
likely progenitors of today's massive spheroidal galaxies.

Given their substantial dust obscuration, the most insightful way to
study SMGs and their star formation properties is through the
dust-reprocessed emission at rest-frame far-infrared (FIR)
wavelengths. While continuum diagnostics at such wavelengths are
particularly useful to determine star formation rates (SFRs)
unaffected by obscuration, the most insightful way to study the fate
of such galaxies is through emission line fluxes, morphology and
dynamics of the material that fuels the star formation, i.e.,
molecular gas (typically CO). Molecular gas was detected in $>$30 SMGs
to date, revealing large gas reservoirs of $>$10$^{10}$\,\msol\ in
most cases (see Solomon \& Vanden Bout \citeyear{sv05} for a
review). However, most of these studies were carried out in mid- to
high-$J$ CO transitions, rather than the fundamental \aco\ transition.

Recent studies of \aco\ emission indicate that (in contrast to
high-$z$ quasar hosts) this line appears to carry a higher brightness
temperature than the mid- to high-$J$ lines in several SMGs,
suggesting relatively low gas excitation (e.g., Hainline et al.\
\citeyear{hai06}; Riechers et al.\ \citeyear{rie06},
\citeyear{rie10}; Carilli et al.\ \citeyear{car10}; Ivison et al.\
\citeyear{ivi10}; Harris et al.\ \citeyear{har10}).  Therefore,
earlier studies may systematically underestimate the total amount of
molecular gas that is present in SMGs.  The low excitation in at least
part of the molecular gas reservoirs raises the question whether the
commonly used $\alpha_{\rm CO}$ conversion factor from CO luminosity
to gas mass ($M_{\rm gas}$) for ultra-luminous infrared galaxies
(ULIRGs) in the nearby universe (Downes \& Solomon \citeyear{ds98}) is
applicable, or if this practice leads to an underprediction of $M_{\rm
gas}$.

\begin{figure*}
\epsscale{1.15}
\vspace{-3mm}

\plotone{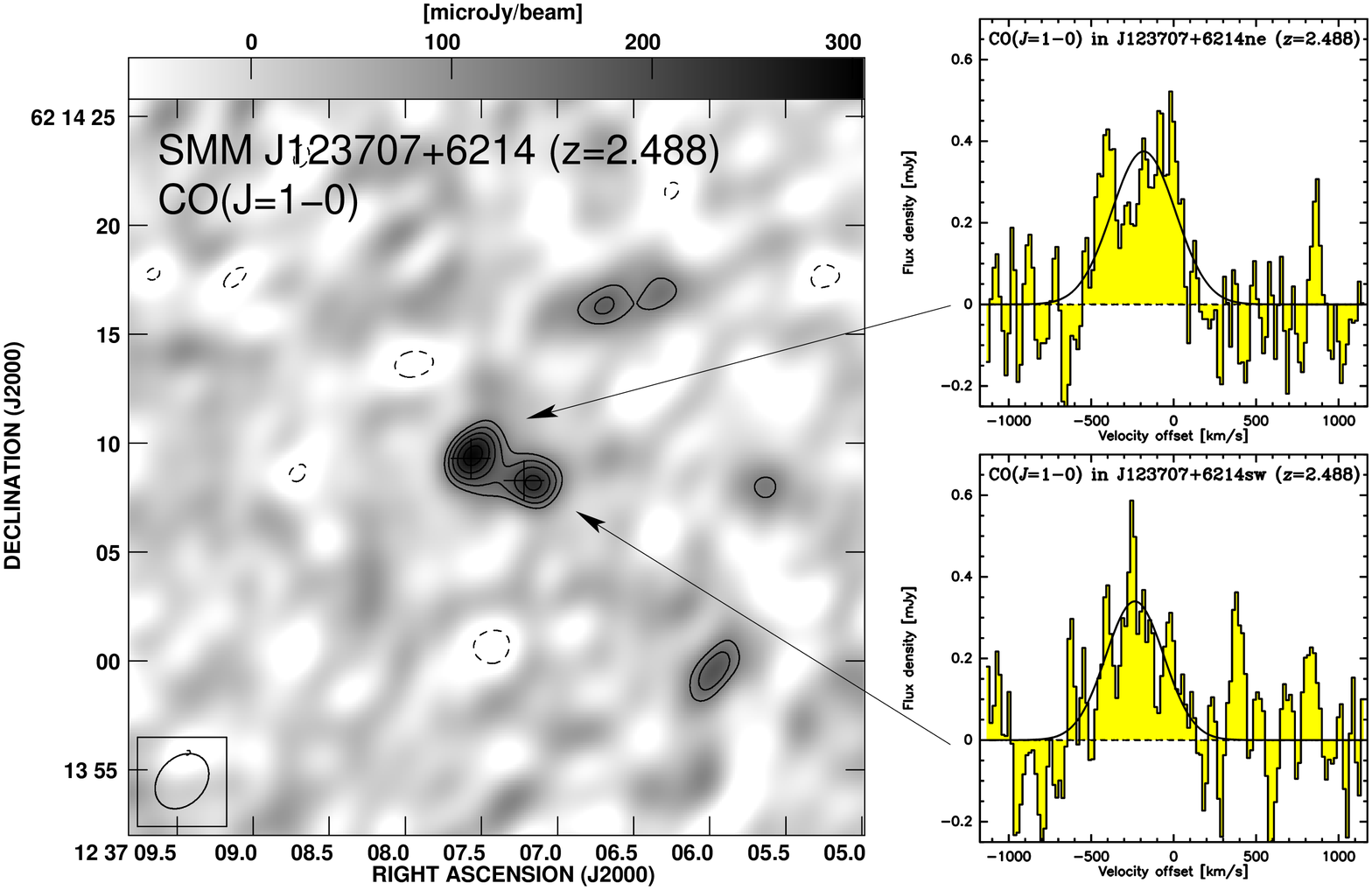}
\vspace{-11mm}

\caption{EVLA \aco\ emission  toward J123707+6214. {\em Left}: 
Velocity-integrated \aco\ map over 490\,\kms\ (54\,MHz). At a
resolution of 2.8$''$$\times$2.1$''$ (as indicated in the bottom
left), the emission is resolved into a north-east (ne) and south-west
(sw) component. The crosses indicate the peak positions of the \cco\
emission (Tacconi et al.\ \citeyear{tac06}). Contours are shown in
steps of 1$\sigma$=39\,$\mu$Jy\,beam$^{-1}$, starting at
$\pm$3$\sigma$. {\em Right}: Beam-corrected \aco\ spectra toward the
ne ({\em top}) and sw ({\em bottom}) components at 18\,\kms\ (2\,MHz)
resolution (histograms), along with Gaussian fits to the line emission
(black curves). The velocity scale is relative to the tuning frequency
of 33.029\,GHz (corresponding to $z$=2.49).
\label{f1}}
%
\end{figure*}

To further investigate this issue, we have started a systematic
Expanded Very Large Array (EVLA) survey of \aco\ emission in high-$z$
SMGs and quasar host galaxies that were previously detected in
higher-$J$ CO lines.  In this Letter, we report the first results from
this study, i.e., the detection of spatially extended \aco\ emission
toward the $z$=2.49 SMG J123707+6214 (GN19; HDF242; Borys et al.\
\citeyear{bor03}). We also report the detection of \eco\ emission in
this system, using the Plateau de Bure Interferometer (PdBI). We use a
concordance, flat $\Lambda$CDM cosmology throughout, with
$H_0$=71\,\kms\,Mpc$^{-1}$, $\Omega_{\rm M}$=0.27, and
$\Omega_{\Lambda}$=0.73 (Spergel \etal\ \citeyear{spe03},
\citeyear{spe07}).

\section{Observations}

\subsection{EVLA}

We observed the \aco\ ($\nu_{\rm rest} = 115.2712$\,GHz) emission line
toward J123707+6214 using the EVLA.  At $z$=2.49, this line is
redshifted to 33.029\,GHz (9.08\,mm).  Observations were carried out
under excellent weather conditions (typical atmospheric phase rms:\
2.3$^\circ$ on a 300\,m baseline) in D array on 2010 April 11 (NRAO
Legacy ID:\ AR708), resulting in 2.9\,hr on-source time with
18\,antennas (equivalent to 1.2\,hr with 27\,antennas) after rejection
of bad data. The nearby ($5.4^\circ$ distance) quasar J1302+5748 was
observed every 7.5\,minutes for pointing, secondary amplitude and
phase calibration. For primary flux calibration, the standard
calibrator 3C286 was observed, leading to a calibration that is
accurate within $\lesssim$10\%. Observations were set up using a
total bandwidth of 252\,MHz (dual polarization; after rejection of
overlapping edge channels between sub-bands; corresponding to
$\sim$2300\,\kms\ at 9.08\,mm) with the WIDAR correlator.

For data reduction and analysis, the AIPS package was used.  All data
were mapped using `natural' weighting.  Maps of the
velocity-integrated CO $J$=1$\to$0 line emission yield a synthesized
clean beam size of 2.8$''$$\times$2.1$''$ at an rms noise level of
39\,$\mu$Jy\,beam$^{-1}$ over 490\,\kms\ (54\,MHz).

\begin{figure*}
\epsscale{1.15}
\vspace{-3mm}

\plotone{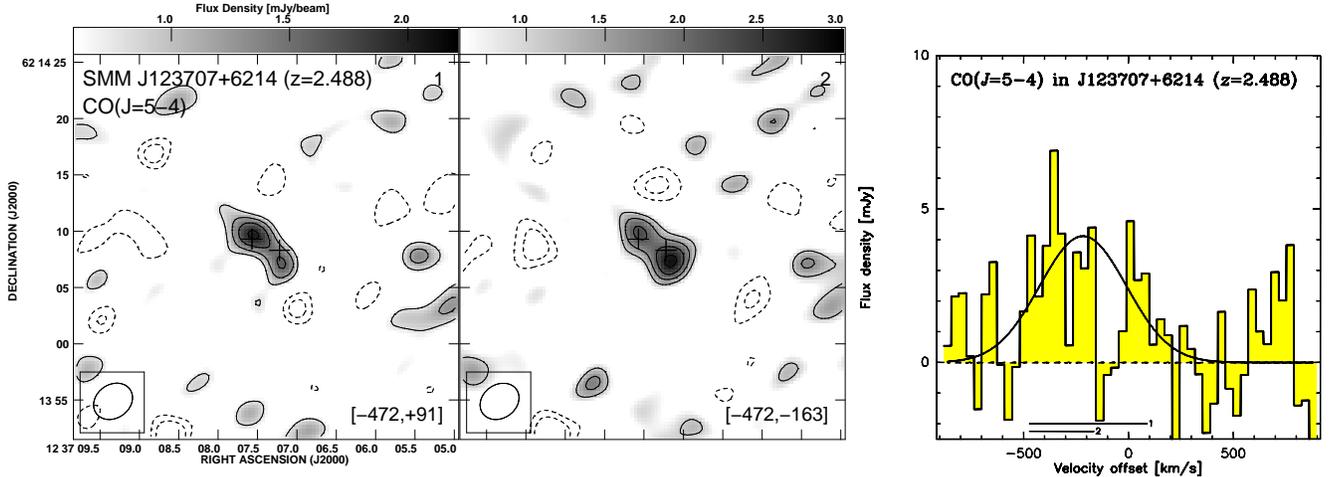}
\vspace{-5mm}

\caption{PdBI \eco\ emission  toward J123707+6214. {\em Left/Middle}: 
Velocity-integrated \eco\ maps over 563/309\,\kms\ (320/180\,MHz). At
a resolution of 3.7$''$$\times$3.0$''$ (as indicated in the bottom
left), the emission is resolved into the ne and sw components also
seen in \aco. The velocity ranges are selected to maximize the
integrated ({\em left}) and sw component ({\em middle})
signal-to-noise ratio.  The crosses are the same as in Fig.~\ref{f1}.
Contours are shown in steps of 1$\sigma$=0.39/0.54\,mJy\,beam$^{-1}$,
starting at $\pm$2$\sigma$. {\em Right}: Integrated \eco\ spectrum at
36\,\kms\ (20\,MHz) resolution (histogram), along with a Gaussian fit
to the line emission (black curve). The velocity ranges of the two
maps are indicated by the black lines. The velocity scale is relative
to the tuning frequency of 165.1198\,GHz (corresponding to $z$=2.49).
\label{f1x}}
%
\end{figure*}

\subsection{PdBI}

We observed the \eco\ ($\nu_{\rm rest} = 576.2679$\,GHz) emission line
toward J123707+6214 using the IRAM PdBI.  At $z$=2.49, this line is
redshifted to 165.1198\,GHz (1.82\,mm).  Observations were carried out
under good weather conditions with five antennas in the compact D
configuration during 3\,tracks on 2008 July 10, and August 09 and 14
(IRAM program ID:\ SC47), for a total of 14.9\,hr, resulting in
8.1\,hr of 6 antenna equivalent on-source time after rejection of bad
data.  The nearby quasars B0954+658 and B1418+546 (distance to
J123707+6214:\ $17.4^\circ$ and $15.4^\circ$) were observed every
20\,minutes for pointing, secondary amplitude and phase
calibration. For primary flux calibration, the standard calibrators
MWC\,349 and 3C454.3 were observed. Observations were set up using a
total spectrometer bandwidth of 1\,GHz (dual polarization;
corresponding to $\sim$1800\,\kms\ at 1.82\,mm).

For data reduction and analysis, the GILDAS package was used.  All
data were mapped using `natural' weighting. Maps of the
velocity-integrated CO $J$=5$\to$4 line emission yield a synthesized
clean beam size of 3.7$''$$\times$3.0$''$ at an rms noise level of
0.39/0.54\,mJy\,beam$^{-1}$ over 563/309\,\kms\ (320/180\,MHz).

\section{Results}

\subsection{Gas Morphology and Emission Line Properties}

We have detected spatially resolved \aco\ line emission toward the
$z$=2.49 SMG J123707+6214, measuring two components at 7$\sigma$
(north-east; `ne')\footnote{Following the nomenclature by Tacconi et
al.\ (\citeyear{tac06}).} and 6$\sigma$ (south-west; `sw')
significance in the velocity-integrated emission line map (Fig.\
\ref{f1}, {\em left}).  We do not detect the underlying continuum
emission at a 3$\sigma$ upper limit of 54\,$\mu$Jy\,beam$^{-1}$ at
9.08\,mm (rest-frame 2.6\,mm).

From Gaussian fitting to the \aco\ line profiles of the ne and sw
components (Fig.\ \ref{f1}, {\em right}),\footnote{\aco\ spectra are
Hanning-smoothed.} we obtain line peak strengths of
$S_{\nu}$=375$\pm$57 and 340$\pm$64\,$\mu$Jy at line FWHMs of
d$v$=454$\pm$87 and 414$\pm$92\,\kms, centered at redshifts of
$z$=2.4879$\pm$0.0005 and 2.4873$\pm$0.0005, respectively. The line
widths and redshifts are consistent with those measured for the \cco\
emission line (Tacconi et al.\ \citeyear{tac06}), and are equal for
both spatial components within the uncertainties. From the
spatially-integrated emission, we determine a systemic redshift of
$z$=2.4876$\pm$0.0004, which we adopt as the nominal value for the
system in the following. The line parameters for J123707+6214ne and sw
correspond to velocity-integrated emission line strengths of $I_{\rm
CO(1-0)}$=0.180$\pm$0.029 and 0.149$\pm$0.029\,Jy\,\kms, i.e., line
luminosities of $L'_{\rm CO(1-0)}$=(5.36$\pm$0.86) and
(4.43$\pm$0.85)$\times$10$^{10}$\,\lprime.

We have also detected spatially resolved \eco\ line emission at
$\gtrsim$5$\sigma$ significance toward both components of J123707+6214
(Fig.\ \ref{f1x}). The ne component dominates the integrated line
emission (Fig.\ \ref{f1x}, {\em left}), extracted over a velocity
range comparable to the \aco\ map. The maximum signal-to-noise ratio
on the sw component is obtained over a narrower velocity range (Fig.\
\ref{f1x}, {\em middle}). In this map, the sw component is brighter than 
the ne component, comparable to what is seen in the \cco\ maps of
Tacconi et al.\ (\citeyear{tac06}, \citeyear{tac08}). Only $\sim$60\%
of the emission from the ne component are seen over this narrower
velocity range. We do not detect the underlying continuum emission at
a 3$\sigma$ upper limit of 0.8\,mJy\,beam$^{-1}$ at 1.82\,mm
(rest-frame 520\,$\mu$m).

From Gaussian fitting to the integrated \eco\ line profile (Fig.\
\ref{f1x}, {\em right}), we obtain $S_{\nu}$=4.1$\pm$0.8\,mJy 
at d$v$=485$\pm$110\,\kms, centered at $z$=2.4875$\pm$0.0006. This
corresponds to $I_{\rm CO(5-4)}$=2.12$\pm$0.51\,Jy\,\kms. 
Fitting the ne and sw components individually yields
$S_{\nu}$=2.4$\pm$0.5 and 2.1$\pm$0.5\,mJy, d$v$=467$\pm$124 and
432$\pm$130\,\kms, and $I_{\rm CO(5-4)}$=1.17$\pm$0.33 and
0.94$\pm$0.29\,Jy\,\kms, respectively.  We thus derive $L'_{\rm
CO(5-4)}$=(1.39$\pm$0.32) and (1.12$\pm$0.29)$\times$10$^{10}$\,\lprime, 
respectively.

\begin{figure*}
\epsscale{1.2}
\vspace{-9mm}

\plotone{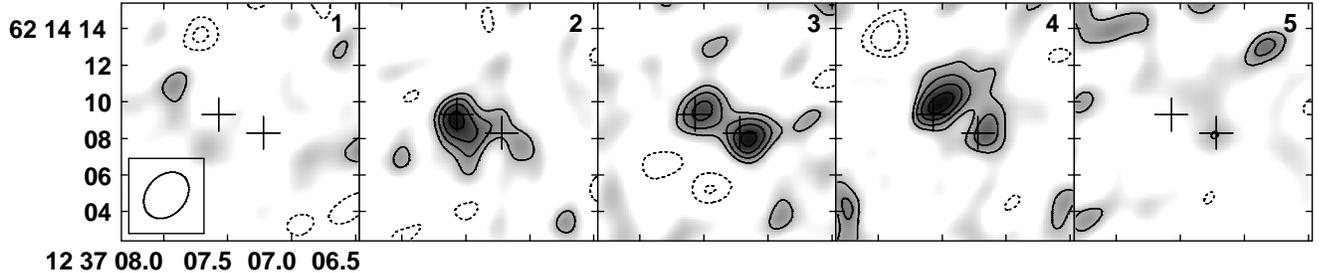}
\vspace{-23mm}

\caption{Channel maps of the \aco\ emission in J123707+6214. One channel 
width is 182\,\kms\ (20\,MHz; frequencies increase with channel number 
and are shown at 33.010, 33.030, 33.050, 33.070, and 33.090\,GHz). 
Contours are shown in steps of 1$\sigma$=57\,$\mu$Jy\,beam$^{-1}$, 
starting at $\pm$2$\sigma$. The beam size (as indicated in the bottom 
left) and crosses are the same as in Fig.~\ref{f1}.
\label{f2}}
%
\end{figure*}

This implies CO $J$=3$\to$2/1$\to$0 line brightness temperature ratios
of $r_{31}$=0.39$\pm$0.09 (ne)\footnote{We recomputed the \cco\ fluxes
based on the data presented in Tacconi et al.\ (\citeyear{tac06}) by
extracting the emission over the same (broader) velocity range as the
\aco\ and \eco\ emission. This yields $I_{\rm CO(3-2)}$=0.63$\pm$0.10 and 
0.50$\pm$0.10\,Jy\,\kms\ for the ne and sw components,
respectively. The different methods for extracting fluxes are likely
responsible for differences between our analysis and an independent
\aco\ study carried out in parallel by Ivison et al.\ (\citeyear{ivi11}).}
and 0.37$\pm$0.10 (sw), and CO $J$=5$\to$4/1$\to$0 line brightness
temperature ratios of $r_{51}$=0.26$\pm$0.07 and 0.25$\pm$0.08. The
\cco\ and \eco\ emission lines are clearly subthermally excited toward
both components ($r_{31}$$<$1 and $r_{51}$$<$1). Interestingly, both
components appear to have comparable gas excitation.  The ne component
is brighter in all CO transitions.  This suggests that the ne
component carries the dominant fraction of the molecular gas mass in
this system.  Based on a ULIRG conversion factor $\alpha_{\rm
CO}$=0.8\,\msol\,(\lprime )$^{-1}$ to derive $M_{\rm gas}$ from
$L'_{\rm CO(1-0)}$ (Downes \& Solomon \citeyear{ds98}), we determine
the total molecular gas masses of J123707+6214ne and sw to be $M_{\rm
gas}$=4.3 and 3.5$\times$10$^{10}$\,\msol,\footnote{A Milky-Way-like
$\alpha_{\rm CO}$=3.5\,\msol\,(\lprime )$^{-1}$ (e.g., Daddi et al.\
\citeyear{dad10}) would increase $M_{\rm gas}$ by a factor of 4.4.}
i.e., by more than a factor of 2 higher than previously found based on
the \cco\ data (scaled to the same $\alpha_{\rm CO}$), and
corresponding to $\sim$2/3 of the stellar mass in this system (Tacconi
et al.\ \citeyear{tac06}, \citeyear{tac08}).

\subsection{Dynamical Structure of the Gas Reservoir}

In Figure~\ref{f2}, maps of the \aco\ emission are shown in 182\,\kms\
wide velocity channels. The emission toward J123707+6214ne appears
dynamically resolved on $\sim$1.5$''$ ($\sim$12\,kpc) scales, which
may suggest that the emission is more spatially extended than in the
\cco\ line (0.5$''$$\pm$0.2$''$, or 4.1$\pm$1.6\,kpc; Tacconi et al.\
\citeyear{tac06}). J123707+6214sw appears marginally spatially resolved in
position-velocity space at best, consistent with the size measured in
\cco\ emission within the relative uncertainties (0.9$''$$\pm$0.3$''$,
or 7.4$\pm$2.5\,kpc; Tacconi et al.\ \citeyear{tac06}).  Assuming
radii of 6 and 3.7\,kpc for J123707+6214ne and sw, this yields
dynamical masses of $M_{\rm dyn}$\,sin$^2$$i$=2.9 and
1.5$\times$10$^{11}$\,\msol\ (which we estimate to be reliable within
a factor of 2). This is about twice as high as previous estimates
based on \cco\ emission (Tacconi et al.\ \citeyear{tac08}), and
corresponds to gas mass fractions of $f_{\rm gas}$=0.15 and 0.23 for
J123707+6214ne and sw, respectively. Higher resolution observations
are required to better constrain how the merger dynamics impact the CO
line profiles and the morphology of the gas reservoir, which is
necessary to determine more precise dynamical masses.

\section{Analysis and Discussion}

\subsection{Origin of the CO Emission}

\begin{figure*}
\vspace{-8mm}

\epsscale{1.15}
\plotone{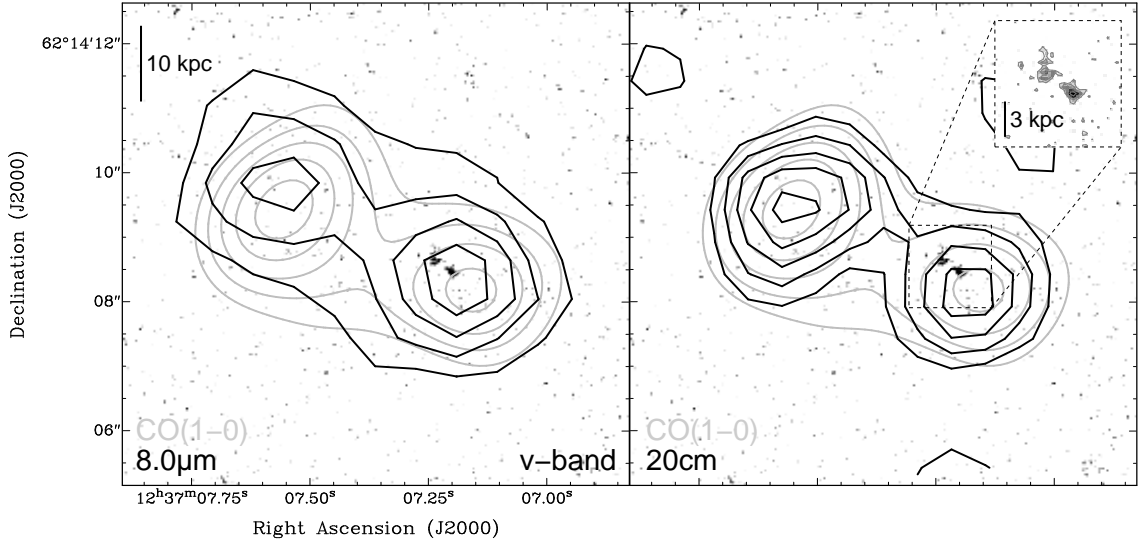}
\vspace{-3mm}

\caption{Overlays of the \aco\ emission (gray contours) on top of
an {\em HST}/ACS $v$-band image (gray scale and inset; from Giavalisco
et al.\ \citeyear{gia04}). In addition, the {\em Spitzer}/IRAC
8.0\,$\mu$m ({\em left}; PSF FWHM = 2$''$; from Dickinson et al., in
prep.) and VLA 20\,cm ({\em right}; 1.7$''$$\times$1.6$''$ resolution;
from Morrison \etal\ \citeyear{mor10}) emission are shown (black
contours in each panel).
\label{f4}}
%
\end{figure*}

SMGs are commonly associated with heavily obscured
starbursts. J123707+6214 is a particularly insightful example of this
population, as the ne component remains undetected at all wavelengths
shortward of 3.6\,$\mu$m (rest-frame 1.0\,$\mu$m). As shown in
Figure~\ref{f4}, the peak of the \aco\ emission of the ne component is
clearly associated with peaks in the mid-infrared (8.0\,$\mu$m;
rest-frame 2.3\,$\mu$m) and radio continuum (20\,cm; rest-frame 6\,cm;
see also Tacconi et al.\ \citeyear{tac08}). The ne component also
slightly dominates the radio emission ($\sim$55\%), which suggests
that it contributes the dominant fraction to the source's SFR.  As
also shown in Fig.~\ref{f4}, J123707+6214sw consists of multiple
components in the optical (606\,nm; rest-frame 174\,nm) that are
separated by a few kpc (see also Swinbank et al.\
\citeyear{swi04}). This may correspond to multiple star-forming
clumps, embedded in a more complex, extended molecular gas reservoir,
but may also reflect the high degree of obscuration in this source.

\subsection{Gas Surface Densities and Star Formation Timescales}

Estimating that the \aco\ emission in J123707+6214ne and sw is
distributed over 6 and 3.7\,kpc radius regions, the surface-averaged
gas densities are $\Sigma_{\rm gas}$=3.8 and
8.1$\times$10$^8$\,\msol\,kpc$^{-2}$, i.e., comparable to but somewhat
lower than estimates based on the (assuming the above size estimates)
more compact \cco\ emission (Tacconi et al.\ \citeyear{tac06},
\citeyear{tac08}). This would be consistent with some of the \aco\
emission being in a diffuse, low surface brightness component.

Based on the SFR of 500$\pm$250\,\msol\,yr$^{-1}$ determined by
Tacconi et al.\ (\citeyear{tac08}), we derive a gas depletion
timescale of $\tau_{\rm dep}$=$M_{\rm gas}$/SFR $\sim$160\,Myr for
J123707+6214. This is consistent with but on the high end of what is
found for other SMGs (for which $M_{\rm gas}$ are inferred from
mid-$J$ CO lines; e.g., Greve \etal\ \citeyear{gre05}).

\section{Conclusions}

We have detected spatially resolved \aco\ and \eco\ emission toward
the $z$=2.49 SMG J123707+6214.  We resolve the emission into two
components previously detected in \cco\ emission (Tacconi et al.\
\citeyear{tac06}, \citeyear{tac08}), which are likely merging
galaxies. Both components show similar CO excitation properties, with
moderate $J$=3$\to$2/1$\to$0 line ratios of $r_{31}$$\sim$0.38, and
relatively low $J$=5$\to$4/1$\to$0 line ratios of
$r_{51}$$\sim$0.25. The implied $J$=5$\to$4/3$\to$2 line ratios of
$r_{53}$$\sim$0.66 are comparable to those found in other SMGs that
show evidence for mergers (e.g., Wei\ss\ et al.\ \citeyear{wei05}). On
the other hand, the low $r_{31}$ are comparable to those found in
massive gas-rich star-forming galaxies with much lower SFRs
(Dannerbauer et al.\ \citeyear{dan09}; Aravena et al.\
\citeyear{ara10}). This may suggest that, in addition to the
highly-excited gas associated with the starburst, J123707+6214 hosts a
substantial amount of low-excitation gas.

The \aco\ emission suggest the presence of $\gtrsim$2.5$\times$ more
molecular gas than expected if assuming a constant brightness
temperature from \cco.  The optically detected merger component (sw)
carries $\sim$45\% of the gas mass in this system, suggesting
comparable amounts of gas in both components, with a slightly higher
contribution coming from the optically invisible component (ne;
$\sim$55\%).  The radio continuum emission consistently indicates a
comparable starburst strength in both components. Assuming that none
is substantially contaminated by an obscured AGN, and given the high
SFR of 500$\pm$250\,\msol\,yr$^{-1}$, this provides supporting
evidence for a ULIRG-like $\alpha_{\rm CO}$ in both components.  The
\aco\ emission in J123707+6214 likely arises from the same gas phase
detected in the higher-$J$ lines, but the \aco\ emission appears
somewhat more spatially extended. This yields a revised,
$\sim$2$\times$ higher estimate for the dynamical mass of the
system. Also, this finding would be consistent with the presence of
some diffuse, low-excitation gas (which may have a higher $\alpha_{\rm
CO}$ than the highly-excited gas). Such a low-excitation component
could be associated with gas that is redistributed by mechanical
energy input from the starburst, or with tidal structure in the
ongoing, gas-rich merger in this system.

Our findings highlight the importance of observing multiple CO lines
including \aco\ to determine the total molecular gas mass and gas
properties in SMGs (as already acknowledged by Tacconi et al.\
\citeyear{tac08} in the initial observations of this source).
Our results are consistent with those found for other SMGs observed in
\aco\ emission (Hainline et al.\ \citeyear{hai06}; Carilli et al.\
\citeyear{car10}; Ivison et al.\ \citeyear{ivi10}; Harris et al.\
\citeyear{har10}), which commonly show lower CO line excitation than
typically found in FIR-luminous quasar host galaxies at comparable
redshifts and with comparable gas masses (e.g., Riechers et al.\
\citeyear{rie06}; \citeyear{rie09}; Wei\ss\ et al.\ \citeyear{wei07}).
This provides supporting evidence that both populations trace
different evolutionary stages of the same massive galaxy population,
as would be expected in the ULIRG-quasar transition scenario proposed
by Sanders et al.\ (\citeyear{san88}).

J123707+6214 is a prototypical example of an SMG during an early
merger stage, found in the peak epoch of galaxy formation. Higher
resolution, dynamical mapping of \aco\ emission in this intriguing
system (and others) is desirable to narrow down $\alpha_{\rm CO}$
through dynamical mass measurements over several resolution elements,
as possible with the full EVLA in the future. A more complete census
of \aco\ observations of SMGs will provide the necessary context to
interpret the results of such investigations. Such studies provide the
most direct means to constrain the gas fraction, total mass and
evolutionary state of SMGs, which is necessary to better understand
the evolutionary path of massive galaxies through their most active
phases, and to constrain the molecular gas mass density of the
universe.

\acknowledgments 
We thank the referee for a critical reading of the manuscript and for
a helpful report. DR acknowledges support from NASA through Hubble
Fellowship grant HST-HF-51235.01 awarded by STScI, operated by AURA
for NASA, under contract NAS\,5-26555.  The EVLA is a facility of
NRAO, operated by AUI, under a cooperative agreement with the NSF.

\end{document}